\documentclass[a4paper]{article}
\usepackage{epsfig}
\usepackage{amsmath}
\usepackage{amsfonts}
\usepackage{amssymb}
\usepackage{graphicx}
\usepackage{bm} 
\usepackage{apalike}
\begin{document}

\title{A nonequilibrium-potential approach to competition in neural populations}
\author{R. R. Deza$^{1,*}$, J. I. Deza$^{1,2}$, N. Mart\'{\i}nez$^1$, J. F. Mejias$^3$, and H. S. Wio$^4$
\\\\
\small{$^{1}$ IFIMAR, Faculty of Exact \& Natural Sci.\ (FCEyN),}\\ 
\small{National Univ.\ of Mar del Plata (UNMdP), Mar del Plata, Argentina}\\
\small{$^{2}$Faculty of Engineering, Universidad Atl\'antida Argentina, Mar del Plata, Argentina}\\
\small{$^{3}$Swammerdam Institute for Life Sciences, Center for Neuroscience,}\\ 
\small{University of Amsterdam, Amsterdam, the Netherlands}\\
\small{$^{4}$IFISC (Institute for Cross-Disciplinary Physics \& Complex Systems,}\\ 
\small{UIB \& CSIC), Univ.\ of the Balearic Islands (UIB), Palma de Mallorca, Spain}
}
\maketitle

\begin{abstract}
Energy landscapes are a useful aid for the understanding of dynamical systems, and a valuable tool for their analysis. For a broad class of rate models of neural networks, we derive a global Lyapunov function which provides an energy landscape without any symmetry constraint. This newly obtained ``nonequilibrium potential'' (NEP) predicts with high accuracy the outcomes of the dynamics in the globally stable cases studied here. Common features of the models in this class are bistability---with implications for working memory and slow neural oscillations---and ``population burst'', also relevant in neuroscience. Instead, limit cycles are not found. Their nonexistence can be proven by resort to the Bendixson--Dulac theorem, at least when the NEP remains positive and in the (also generic) singular limit of these models. Hopefully, this NEP will help understand average neural network dynamics from a more formal standpoint, and will also be of help in the description of large heterogeneous neural networks.
\end{abstract}

\section{INTRODUCTION}\label{sec:1}
The analysis of dissipative, autonomous\footnote{The framework can also be applied to nonautonomous flows, as far as their explicit time-dependence can be regarded as slow in comparison with the relaxation times toward the system's attractors (adiabatic approximation).} dynamic flows (especially high-dimensional ones) can be greatly simplified, if a function can be constructed to provide an ``energy landscape'' to the problem. Energy landscapes not only help visualize the systems' phase space and its structural changes as parameters are varied, but allow predicting the rates of activated processes \cite{Wio,epjst146:111,wdl}. Some fields that benefit from the energy landscape approach are optimization problems \cite{kiea83}, neural networks \cite{Yan2013}, protein folding \cite{waea12}, cell nets \cite{waea06}, gene regulatory networks \cite{kiwa07,waea10}, ecology \cite{liea11}, and evolution \cite{zhea12}.

For continuous-time flows, the possibility of this ``Lyapunov function''---with its distinctive property \(\dot{\Phi}<0\) outside the attractors---was suggested in the context of the general stability problem of dynamical systems \cite{lyap} and in a sense, it adds a quantitative dimension to the qualitative theory of differential equations. The linearization of the flow around its attractors always provides such a function, but its validity breaks down well inside their own basin. Instead, finding a \emph{global} Lyapunov function is not an easy problem. If only the information of the (deterministic) dynamical system is to be used, this function can be found for the so-called ``gradient flows''---purely irrotational flows in 3d, exact (longitudinal) forms in any dimensionality. But since for general relaxational flows (having nontrivial Helmholtz--Hodge decomposition) the integrability conditions are not automatically met, some more information is needed.

A hint of what information is needed comes from recalling that dynamical systems are \emph{models} and as such, they leave aside a multitude of degrees of freedom---deemed irrelevant to the model, but nonetheless coupled to the ``system''. A useful framework to deal with them is the one set forward by Langevin \cite{lang}, which makes the dynamical flow into a stochastic process (thus nonautonomous, albeit driven by a stationary ``white noise'' process).

What Graham and his collaborators \cite{graham,grte90} realized more than thirty years ago is that \emph{even in the deterministic limit}, this space enlargement can eventually help meet the integrability conditions. Given an initial state \(x_i\) of a (continuous-time, dissipative, autonomous) dynamic flow \(\dot{x}=f(x)\), its conditional probability density function (pdf) \(P(x,t|x_i,0)\) when submitted to a (Gaussian, centered) white noise \(\xi(t)\) with variance \(\gamma\), namely\footnote{We follow the usual notation in physics, which strictly means \(dx=f(x)dt+dW(t)\) in terms of the Wiener process \(W(t)=\int_0^t\xi(s)ds\). Note that \(dx-dW(t)\) is still a (deterministic) dynamic flow, which implies some kind of connection in the \(x-W\) bundle (but not the one of gauge theory).}
\[\dot{x}=f(x)+\xi(t)\mbox{, with }\langle\xi(t)\rangle=0\mbox{ and }\langle\xi(t)\,\xi(t')\rangle=2\gamma\,\delta(t-t')\]
obeys the Fokker--Planck equation (FPE)
\[\partial_tP(x,t|x_i,0)+\partial_xJ(x,t|x_i,0)=0\mbox{, with }J(x,t|x_i,0)=D^{(1)}(x)P-\partial_x\left[D^{(2)}(x)P\right]\]
in terms of the ``drift'' \(D^{(1)}=f(x)\) and ``diffusion'' \(D^{(2)}=\gamma\) Kramers--Moyal coefficients \cite{risken,gard04,vaka90}. Being the flow nonautonomous but dissipative, one can expect generically situations of statistical energy balance in which the pdf becomes stationary, \(\partial_tP_\mathrm{st}(x)=0\), thus independent of the initial state. Then by defining \(\Phi(x):=-\int_{x_0}^xf(y)dy\), it is immediate to find \(P_\mathrm{st}(x)=\mathcal{N}(x_0)\exp[-\Phi(x)/\gamma]\).

For an \(n\)--component dynamic flow submitted to \(m\le n\) (Gaussian, uncorrelated, centered) white noises \(\xi_i(t)\) with common variance \(\gamma\),
\[\dot{\mathbf{x}}=\mathbf{f}(\mathbf{x})+\sigma\,\Xi(t),\quad\langle\Xi(t)\rangle=0,\quad\langle\Xi(t)\,\Xi^T(t')\rangle=2\gamma I\,\delta(t-t')\]
(\(\sigma\) is an \(n\times m\) constant matrix) the \emph{nonequilibrium potential} (NEP) has been thus defined \cite{graham} as
\begin{equation}\label{eq:1}
\Phi(\mathbf{x})=-\lim_{\gamma\to0}\gamma\,\ln P_\mathrm{st}(\mathbf{x};\gamma).
\end{equation}
The stationary \(n\)--variable FPE
\[\nabla\cdot[\mathbf{f}(\mathbf{x})P_\mathrm{st}(\mathbf{x})-\gamma\,Q\,\nabla P_\mathrm{st}(\mathbf{x})]=0\]
(\(Q:=\sigma\sigma^\mathrm{T}=Q^\mathrm{T}\)) reduces in this limit to
\begin{equation}\label{eq:2}
\mathbf{f}^\mathrm{T}(\mathbf{x})\nabla\Phi+(\nabla\Phi)^\mathrm{T}Q\,\nabla\Phi=0,
\end{equation}
from which \(\Phi(\mathbf{x})\) can in principle be found. In an attractor's basin, asymptotic stability imposes \(D:=\det Q>0\). In fact, for \(m=n\) (restriction adopted hereafter) it is \(D=(\det\sigma)^2\), which in turn requires \(\sigma\) to be nonsingular. Using Eq.\ (\ref{eq:2}),
\[\dot{\Phi}=\dot{\mathbf{x}}^\mathrm{T}\nabla\Phi=\mathbf{f}^\mathrm{T}(\mathbf{x})\nabla\Phi=-(\nabla\Phi)^\mathrm{T}Q\,\nabla\Phi<0\]
for \(\gamma\to0\). Hence, \(\Phi(\mathbf{x})\) is a \emph{Lyapunov function} for the \emph{deterministic} dynamics.

Equation (\ref{eq:2}) has the structure of a Hamilton--Jacobi equation. This trouble can be circumvented if we can decompose \(\mathbf{f}(\mathbf{x})=\mathbf{d}(\mathbf{x})+\mathbf{r}(\mathbf{x})\), with \(\mathbf{d}(\mathbf{x}):=-Q\,\nabla\Phi\) the \emph{dissipative} part of \(\mathbf{f}(\mathbf{x})\).\footnote{Had we written \(\dot{x}=f(x)+\sigma\xi(t)\) for \(n=1\), then \(\Phi(x):=-\sigma^{-2}\int_{x_0}^xf(y)dy\).} Then Eq.\ (\ref{eq:2}) reads \(\mathbf{r}^\mathrm{T}(\mathbf{x})\nabla\Phi=0\), and \(\mathbf{r}(\mathbf{x})\) is the \emph{conservative} part of \(\mathbf{f}(\mathbf{x})\). Note that \(\mathbf{d}(\mathbf{x})\) is still irrotational (in the sense of the Helmholtz decomposition) but is \emph{not} an exact form (the Hodge decomposition is made in the \emph{enlarged} space).

For \(n=2\) we may write \(\mathbf{r}(\mathbf{x})=\kappa\,\Omega\,\nabla\Phi\), with \(\Omega\) the \(N=1\) symplectic matrix. Hence \(\mathbf{f}(\mathbf{x})=-(Q-\kappa\,\Omega)\nabla\Phi\), with \(\det(Q-\kappa\,\Omega)=D+\kappa^2>0\), and thus
\begin{equation}\label{eq:3}
\nabla\Phi=-(Q-\kappa\,\Omega)^{-1}\mathbf{f}(\mathbf{x}).
\end{equation}
For arbitrary real \(\sigma_{ij}\) we can parameterize
\[\sigma=\left(\begin{array}{cc}\sqrt{\lambda_1}\cos\alpha_1&\sqrt{\lambda_1}\sin\alpha_1\\\sqrt{\lambda_2}\cos\alpha_2&\sqrt{\lambda_2}\sin\alpha_2\end{array}\right)\]
and define
\(\lambda:=\sqrt{\lambda_1\lambda_2}\cos(\alpha_1-\alpha_2)\)
(note that the condition \(D>0\) imposes \(\alpha_2\neq\alpha_1\)). Then
\[Q-\kappa\,\Omega=\left(\begin{array}{cc}\lambda_1&\lambda-\kappa\\\lambda+\kappa&\lambda_2\end{array}\right),\]
and Eq.\ (\ref{eq:3}) reads
\[\nabla\Phi:=\left(\begin{array}{c}\partial_1\Phi\\\partial_2\Phi\end{array}\right)=-\frac1{\det(Q-\kappa\,\Omega)}\left(\begin{array}{c}\lambda_2f_1(\mathbf{x})-(\lambda-\kappa)f_2(\mathbf{x})\\-(\lambda+\kappa)f_1(\mathbf{x})+\lambda_1f_2(\mathbf{x})\end{array}\right)\]
(\(\partial_k\) is a shorthand for \(\partial/\partial x_k\)). If a set \(\{\lambda_1,\lambda_2,\lambda,\kappa\}\) can be found such that \(\Phi(\mathbf{x})\) fulfills the \emph{integrability condition} \(\partial_2\partial_1\Phi=\partial_1\partial_2\Phi\), then a NEP exists. Early successful examples are the complex Ginzburg--Landau equation (CGLE) \cite{degr,moea} and the FitzHugh--Nagumo (FHN) model \cite{wio98,wio99} \footnote{The knowledge of a NEP for FHN units has greatly simplified the dynamical analysis of reaction--diffusion \cite{Wio,epjst146:111,wdl} and network \cite{Deza2009,Izus2010,Deza2014} FHN models, as well as the study of stochastic resonance in some extended systems \cite{epjst146:111,wvhb2002}.}. This scheme has been later reformulated \cite{ao04}, extended \cite{wuwa13}, and exploited in many interesting cases \cite{Yan2013,waea12,waea06,kiwa07,waea10,liea11,zhea12}.

The goal of this work is to show that a NEP exists for a broad class of rate models of neural networks, of the type proposed by Wilson and Cowan \cite{cowan72}. Section \ref{sec:2} is devoted to an analysis of the model and variations of. Section \ref{sec:3}, to the derivation of the NEP in some of the cases studied in section \ref{sec:2}. Section \ref{sec:4}, to thorough discussion of our findings. Section \ref{sec:5} collects our conclusions.
\section{THE WILSON--COWAN MODEL}\label{sec:2}
Elucidating the architecture and dynamics of the neocortex is of utmost importance in neuroscience. But despite ongoing titanic efforts like the Human Brain Project or the BRAIN initiative, we are still very far from that goal. Given that the dynamics of single typical neurons has been relatively well described (in some cases even by analytical means), a fundamental approach can be practiced for small neural circuits. This means describing them as networks of excitable elements (neurons) connected by links (synapses), and solving the network dynamics by hybrid (analytical-numerical) techniques. However, the time employed in the analytical solution has poor scaling with size. Hence, this approach becomes unworkable for large (\(>100\)) networks of recurrently interconnected neurons, and one has to rely only on numerical simulations.

Fortunately---as evidenced since long ago by the existence (as in a medium) of wavelike excitations---the huge connectivity of the neocortex enables \emph{coarse-grained} or mean-field descriptions, which provide more concise and relevant information to understand the mesoscopic dynamics of the system. Frequently obtained via mean-field techniques and commonly referred to as \emph{rate models} or \emph{neural mass models}, coarse-grained reductions have been widely used in the theoretical study of neural systems \cite{cowan72,Amit1997,brunel00,wong2006}. In particular, the one proposed by Wilson and Cowan \cite{cowan72} has proved to be very useful in describing the macroscopic dynamics of neural circuits. This level of description is able to capture many of the dynamical features associated with several cognitive and behavioral phenomena, such as working memory \cite{Amit1997,wang2001} or perceptual decision making \cite{wang2002,wong2006}. It is also possible to use a rate model approach to study the dynamics of networks constituted by heterogeneous neurons \cite{Mejias2012,Mejias2014b}, thus recovering part of the complexity lost in the averaging. Disposing of an ``energy function'' (not restricted to symmetric couplings) for rate-level dynamics of neural networks would be a major added advantage.

The Wilson--Cowan model describes the evolution of competing populations \(x_1\), \(x_2\) of \emph{excitatory} and \emph{inhibitory} neurons respectively. The model is defined by \cite{cowan72}
\begin{equation}\label{eq:4}
\tau_1\,\dot{x}_1=-x_1+(\nu_1-r_1x_1)s_1(i_1),\quad\tau_2\,\dot{x}_2=-x_2+(\nu_2-r_2x_2)s_2(i_2),\end{equation}
where \(x_1\) and \(x_2\) represent the coarse-grained activity of an excitatory and an inhibitory neural population, respectively, and the monotonically increasing (sigmoidal) \emph{response functions}
\begin{equation}\label{eq:5}
s_k(i_k):=\frac1{1+\exp[-\beta_k(i_k-i_k^0)]}-\frac1{1+\exp(\beta_ki_k^0)}
\end{equation}
are such that \(s_k(0)=0\), and range from \(-[1+\exp(\beta_ki_k^0)]^{-1}\) for \(i_k\to-\infty\) to \(1-[1+\exp(\beta_ki_k^0)]^{-1}\) for \(i_k\to\infty\). So the first crucial observation about the model is that it is \emph{asymptotically linear}.

The currents \(i_k\) are in turn linearly related to the \(x_k\):
\[\mathbf{i}(\mathbf{x}):=J\mathbf{x}+\mathrm{M}=\left(\begin{array}{cc}j_{11}&-j_{12}\\j_{21}&-j_{22}\end{array}\right)\left(\begin{array}{c}x_1\\x_2\end{array}\right)+\left(\begin{array}{c}\mu_1\\\mu_2\end{array}\right).\]
All the parameters are real and moreover, the \(j_{kl}\) are positive (\(j_{11}\) and \(j_{22}\) are recurrent interactions, \(j_{12}\) and \(j_{21}\) are cross-population interactions). The above definitions are such that for \(\mathrm{M}=0\), \(\mathbf{x}=0\) is a stable fixed point. To avoid confusions in the following, note that \(\det J=-(j_{11}j_{22}-j_{12}j_{21})\).

Wilson and Cowan \cite{cowan72} found interesting features as e.g.\ staircases of bistable regimes and limit cycles. A thorough analysis of the model's bifurcation structure has been undertaken in \cite{boki1992}. The authors create a two-parameter structural portrait by fixing all the parameters but \(\mu_1\) and \(j_{21}\) and find that the \(\mu_1-j_{21}\) plane turns out to be partitioned into several regions by:
\begin{itemize}
\item a fold point bifurcation curve (the number of fixed points changes by two when crossed),
\item an Andronov--Hopf bifurcation curve (separates regions with stable and unstable foci),
\item a saddle separatrix loop (a limit cycle on one side, none on the other), and
\item a double limit cycle curve (the number of limit cycles changes by two).
\end{itemize}

The \emph{uncoupled} case (\(j_{12}=j_{21}=0\)) is clearly a gradient system, with potential
\[\Phi(\mathbf{x})=\frac1{\tau_1}\left\{\frac{x_1^2}2-\frac1{j_{11}}\left[F_1( i_1)-F_1( \mu_1)\right]\right\}+\frac1{\tau_2}\left\{\frac{x_2^2}2-\frac1{j_{22}}\left[F_2( i_2)-F_2( \mu_2)\right]\right\},\]
where \(i_1=j_{11}\,x_1+\mu_1\) and \(i_2=-j_{22}\,x_2+\mu_2\). Functions \(F_k\) differ only in the values of their parameters. Their functional expression, involving polylogs, is uninteresting besides being complicated. Much more insight is obtained by observing the global features:
\begin{itemize}
\item for \(i_k\to-\infty\), Eqs.\ (\ref{eq:4}) become \(\tau_k\,\dot{x}_k=-x_k-(\nu_k-r_kx_k)[1+\exp(\beta_ki_k^0)]^{-1}\),
\item for \(i_k\to\infty\), \(\tau_k\,\dot{x}_k=-x_k+(\nu_k-r_kx_k)(1-[1+\exp(\beta_ki_k^0)]^{-1})\).
\end{itemize}
So it seems interesting to look at them in the limit \(\beta_k\to\infty\) (\(k=1,2\)),
\[\tau_1\,\dot{x}_1=-x_1+(\nu_1-r_1x_1)\theta(i_1-i_1^0),\quad\tau_2\,\dot{x}_2=-x_2+(\nu_2-r_2x_2)\theta(i_2-i_2^0),\]
\(\theta(x)\) being Heaviside's unit step function. Unfortunately, neither Eqs.\ (\ref{eq:4}) nor their singular limit fulfill the above mentioned integrability condition.

In practice however, the names of Wilson and Cowan are associated to the broader class of rate models. In the following we shall show that the model defined by
\begin{equation}\label{eq:6}
\tau_1\,\dot{x}_1=-x_1+s_1(i_1),\quad\tau_2\,\dot{x}_2=-x_2+s_2(i_2)
\end{equation}
does admit a NEP---for \emph{any} functional forms of the nonlinear single-variable functions \(s_k(i_k)\) \footnote{With our mind in neurophysiology applications, we shall assume \(s_k(i_k)\) to have the same functional form, of sigmoidal shape. But neither condition is necessary to satisfy the integrability condition.}---provided global stability is assured.
\section{NONEQUILIBRIUM POTENTIAL}\label{sec:3}
For the model defined by Eq.\ (\ref{eq:6}), it is \(\mathbf{f}(\mathbf{x})=\left(\begin{array}{c}\frac1{\tau_1}[-x_1+s_1(i_1)]\\\frac1{\tau_2}[-x_2+s_2(i_2)]\end{array}\right)\). The condition \(\partial_2\partial_1\Phi=\partial_1\partial_2\Phi\), namely
\[\frac{\lambda_2}{\tau_1}\left[-j_{12}\,s_1^\prime(i_1)\right]-\frac{\lambda-\kappa}{\tau_2}\left[-1-j_{22}\,s_2^\prime(i_2)\right]=-\frac{\lambda+\kappa}{\tau_1}\left[-1+j_{11}\,s_1^\prime(i_1)\right]+\frac{\lambda_1}{\tau_2}\left[j_{21}\,s_2^\prime(i_2)\right],\]
boils down to
\[j_{12}\,\lambda_2=j_{11}\,(\lambda+\kappa),\quad j_{22}\,(\lambda-\kappa)=j_{21}\,\lambda_1,\quad\frac{\lambda-\kappa}{\tau_2}=\frac{\lambda+\kappa}{\tau_1}\]
(and these in turn to \(j_{21}\,j_{11}\,\tau_1\,\lambda_1=j_{12}\,j_{22}\,\tau_2\,\lambda_2\)) so that \(\lambda_2\), \(\lambda\) and \(\kappa\) can be expressed in terms of \(\lambda_1\), which sets the global scale of \(\Phi(\mathbf{x})\):
\[\lambda_2=\frac{j_{21}}{j_{22}}\frac{j_{11}}{j_{12}}\frac{\tau_1}{\tau_2}\,\lambda_1,\quad\lambda=\frac{j_{21}}{j_{22}}\,\frac{\tau_1+\tau_2}{2\tau_2}\,\lambda_1,\quad\kappa=\frac{j_{21}}{j_{22}}\,\frac{\tau_1-\tau_2}{2\tau_2}\,\lambda_1\]
(note that \(\tau_1=\tau_2\) suffices to render the flow \emph{purely dissipative}, albeit not gradient). From this, a good choice is \(\lambda_1:=\frac{j_{22}}{j_{21}}\,\tau_2\rho\). In summary,
\[Q-\kappa\,\Omega=\rho\left(\begin{array}{cc}\frac{j_{22}}{j_{21}}\,\tau_2&\tau_2\\\tau_1&\frac{j_{11}}{j_{12}}\,\tau_1\end{array}\right),\quad\det(Q-\kappa\,\Omega)=-\frac{\rho^2\tau_1\tau_2}{j_{12}j_{21}}\det J,\]
and Eq.\ (\ref{eq:3}) becomes
\begin{equation}\label{eq:7}
\nabla\Phi=\frac{j_{12}j_{21}}{\rho\tau_1\tau_2\det J}\left(\begin{array}{c}\frac{j_{11}}{j_{12}}[-x_1+s_1(i_1)]-[-x_2+s_2(i_2)]\\\,-[-x_1+s_1(i_1)]+\frac{j_{22}}{j_{21}}[-x_2+s_2(i_2)]\end{array}\right).
\end{equation}
Integrating Eq.\ (\ref{eq:7}) over any path from \(\mathbf{x}=0\), yields
\begin{equation}\label{eq:8}
\Phi(\mathbf{x})=-\frac{j_{11}j_{21}x_1^2-2j_{12}j_{21}x_1x_2+j_{12}j_{22}x_2^2}{2\rho\tau_1\tau_2\det J}+\frac{j_{21}\left[S_1( i_1)-S_1(\mu_1)\right]-j_{12}\left[S_2( i_2)-S_2(\mu_2)\right]}{\rho\tau_1\tau_2\det J}.
\end{equation}
\section{THEORETICAL ANALYSIS}\label{sec:4}
The first---crucial---observation is that being \(s_k(i_k)\) sigmoidal functions, Eq.\ (\ref{eq:8}) is \emph{at most quadratic}. Global stability thus imposes \(\det J<0\), i.e.\ \(j_{11}j_{22}>j_{12}j_{21}\). But note that matrix \(J\) also determines the paraboloid's cross section \footnote{Incidentally, note that \(j_{11}j_{21}x_1^2-2j_{12}j_{21}x_1x_2+j_{12}j_{22}x_2^2\equiv(J\mathbf{x})_1(J\mathbf{x})_2-x_1x_2\det J\).}.

If the form \(\mathbf{x}^TA\,\mathbf{x}\), \(A=\left(\begin{array}{cc}j_{11}j_{21}&-j_{12}j_{21}\\-j_{12}j_{21}&j_{12}j_{22}\end{array}\right)\) in the first term of Eq.\ (\ref{eq:8}) could be straightforwardly factored out, then one could tell what its section is by watching whether the factors are real or complex.  A more systematic approach is to reduce \(\mathbf{x}^TA\,\mathbf{x}\) to canonical form by a similarity transformation that involves the normalized eigenvectors of \(A\). Then the inverse squared lengths of the principal axes are the eigenvalues \(\lambda_{1,2}=\frac1{2}(j_{11}j_{21}+j_{12}j_{22})\pm\sqrt{\frac1{4}(j_{11}j_{21}+j_{12}j_{22})^2+j_{12}j_{21}\det J}\) of \(A\). Since the \(j_{kl}\) are positive and \(\det J<0\), the second term is lesser than the first and the cross section is definitely \emph{elliptic}. Although global stability rules out \(\det J>0\), we can conclude that the instability proceeds through a pitchfork (codimension one) bifurcation along the \emph{minor} principal axis (because of the double role of \(\det J\)), not a Hopf (codimension two) one.

For the remaining terms, we note that Eq.\ (\ref{eq:7}) can be written as \(\nabla\Phi=\frac1{\rho\tau_1\tau_2\det J}\left(\begin{array}{cc}j_{21}&0\\0&-j_{12}\end{array}\right)J(\mathbf{x}-\mathbf{s})\) and recall that \(s_k(i_k)\) have sigmoidal shape. So at large \(|\mathbf{x}|\), the component \(-\frac1{\rho\tau_1\tau_2\det J}\left(\begin{array}{cc}j_{21}&0\\0&-j_{12}\end{array}\right)J\mathbf{s}\) will tend to different constants---according to the signs of \(i_k\)---so the asymptotic contribution of these terms will be piecewise linear, namely a collection of half planes.

The reduction to the uncoupled case can be safely done by writing \(j_{12}=\epsilon\) and \(j_{21}=\alpha\epsilon\):
\[\Phi(\mathbf{x})=\frac{j_{11}\alpha\epsilon x_1^2-2\alpha\epsilon^2x_1x_2+j_{22}\epsilon x_2^2}{2\rho\tau_1\tau_2(j_{11}j_{22}-\alpha\epsilon^2)}-\frac{\alpha\epsilon\left[S_1( i_1)-S_1(\mu_1)\right]-\epsilon\left[S_2( i_2)-S_2(\mu_2)\right]}{\rho\tau_1\tau_2(j_{11}j_{22}-\alpha\epsilon^2)}.\]

To first order as \(\epsilon\to0\), one retrieves
\[\Phi(\mathbf{x})\approx\frac{\epsilon}{\rho\tau_1j_{11}}\left\{\frac{\alpha j_{11}}{\tau_2j_{22}}\left[\frac{x_1^2}2-\frac1{j_{11}}\left[S_1( i_1)-S_1( \mu_1)\right]\right]+\frac1{\tau_2}\left[\frac{x_2^2}2-\frac1{j_{21}}\left[S_2( i_2)-S_2( \mu_2)\right]\right]\right\},\]
so by choosing \(\alpha=\frac{\tau_2j_{22}}{\tau_1j_{11}}\) and \(\rho=\frac{\epsilon}{\tau_1j_{11}}\),
\[
\Phi(\mathbf{x})\approx\frac1{\tau_1}\left\{\frac{x_1^2}2-\frac1{j_{11}}\left[S_1( i_1)-S_1( \mu_1)\right]\right\}+\frac1{\tau_2}\left\{\frac{x_2^2}2-\frac1{j_{21}}\left[S_2( i_2)-S_2( \mu_2)\right]\right\}.
\]

A popular choice---that can be cast in the form of Eq.\ (\ref{eq:5})---is \(s_k(i_k):=\frac{\nu_k}2\,(1+\tanh\beta_k i_k)\), \(\beta_k>0\), for which \footnote{Using \(s_k(i_k):=\tanh i_k\), Tsodyks \emph{et al.} have reported a paradoxical increase in \(x_1\) as a result of an increase in \(\mu_2\). Unfortunately, this occurs for \(\det J>0\). What we can assure is that there is a saddle point involved.}
\begin{equation}\label{eq:9}
S_k( i_k)-S_k(\mu_k)=\frac{\nu_k}2\left[i_k-\mu_k+\beta_k^{-1}\ln\frac{\cosh\beta_ki_k}{\cosh\beta_k\mu_k}\right].
\end{equation}
Its \(\beta_k\to\infty\) limit, \(\nu_k\,\theta(i_k)\) with
\begin{equation}\label{eq:10}
S_k( i_k)-S_k(\mu_k)=\nu_k\left[i_k\,\theta(i_k)-\mu_k\,\theta(\mu_k)\right],
\end{equation}
highlights the cores of the response functions while keeping the global landscape \footnote{For \(\mu_k\neq0\), Eq.\ (\ref{eq:10}) can be arrived at from Eq.\ (\ref{eq:9}) given that for \(\beta_k\to\infty\), \(\beta_k^{-1}\ln\cosh\beta_kx\to|x|-\ln 2\). Once Eq.\ (\ref{eq:10}) is obtained, one can let \(\mu_k\to0\).}.

\begin{figure}[htbp]
\begin{center}
\includegraphics[width=.45\columnwidth]{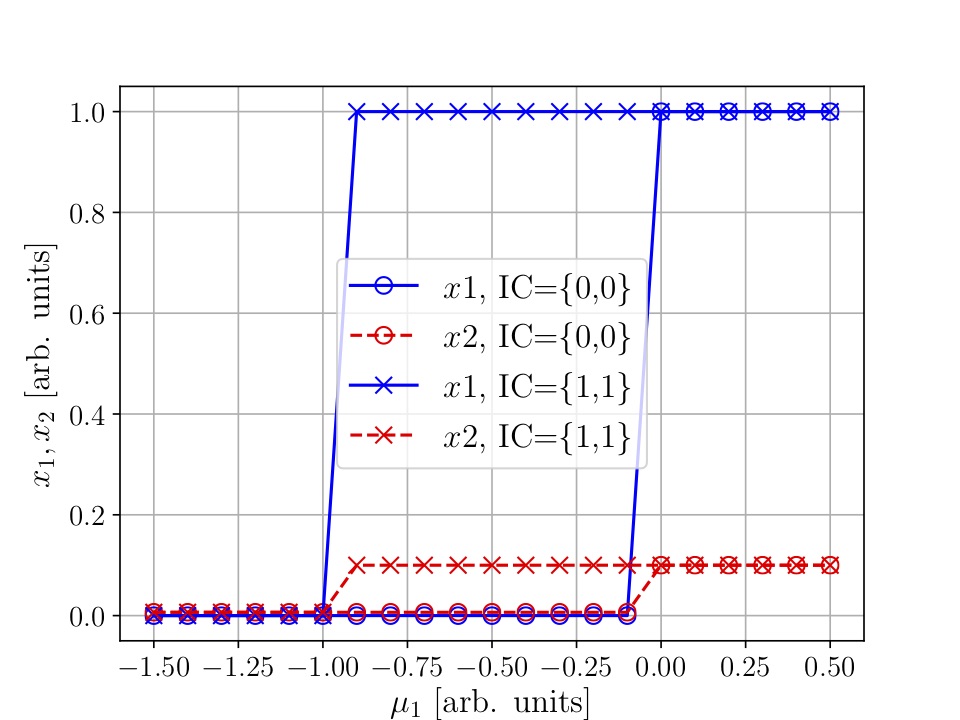}(a)
\includegraphics[width=.45\columnwidth]{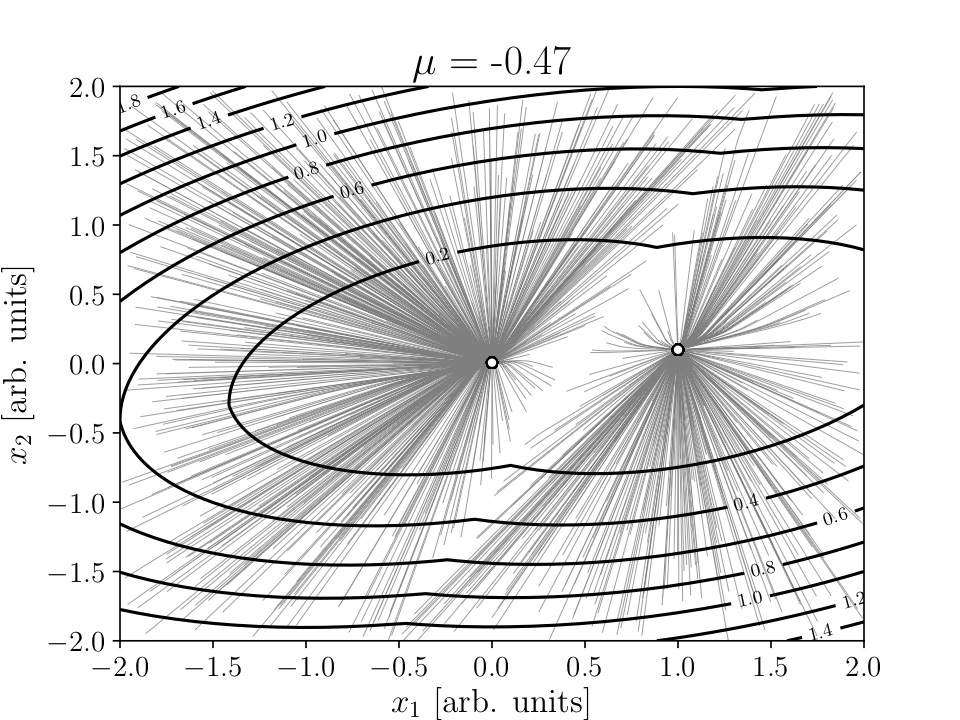}(b)
\includegraphics[width=.45\columnwidth]{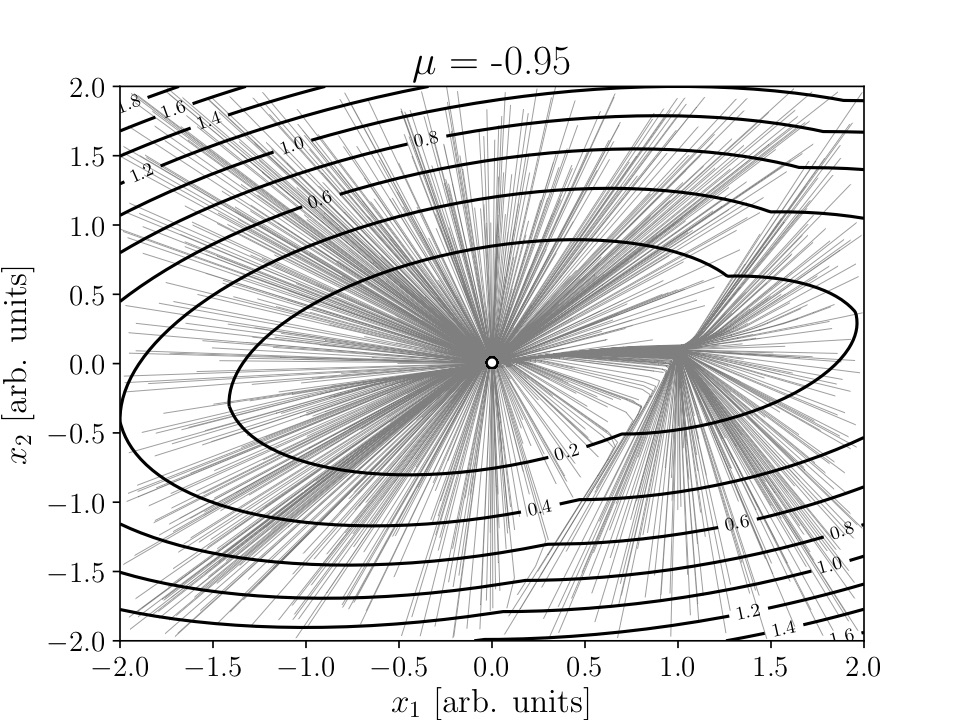}(c)
\includegraphics[width=.45\columnwidth]{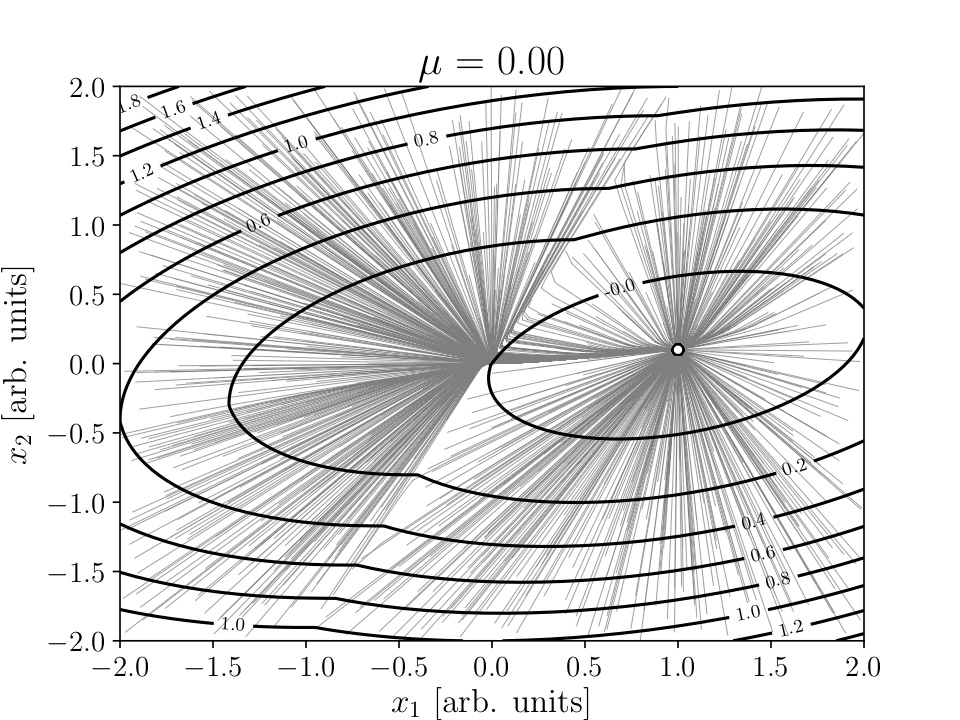}(d)
\caption{Illustration of the analytical results in subsection \ref{ssec:4.1}, for \(\rho=1\), \(\tau_1=\tau_2=1\), \(\mu_2=-.01\), \(\nu_1=1\), \(\nu_2=.1\), \(j_{11}=1\), \(j_{22}=.5\), \(j_{12}=.5\), \(j_{21}=.1\). (a) Abscissas (solid line) and ordinates (dashed line) of the ``off'' (circle) and ``on'' (cross) nodes of Eqs.\ (\ref{eq:6}) with \(s_k(i_k):=\nu_k\,\theta(i_k)\), as functions of \(\mu_1\). Trajectories from random initial conditions and contour plot of the NEP---Eq.\ (\ref{eq:8}), with \(S_k( i_k)-S_k(\mu_k)\) given by Eq.\ (\ref{eq:10})---in the equistable case given by Eq.\ (\ref{eq:11}) (b), and near the ``off'' (c) and ``on'' (d) saddle-node bifurcations.}
\label{fig:1}
\end{center}
\end{figure}

As a check of Eq.\ (\ref{eq:8}), we show in the next subsections the mechanism whereby Eqs.\ (\ref{eq:6}) can sustain bistability.
\subsection{Analytically, for steplike response function \(s_k(i_k):=\nu_k\,\theta(i_k)\)}\label{ssec:4.1}
\begin{itemize}
\item For \(\mu_k<0\) (\(k=1,2\)), there is no question that \(\mathbf{x}=0\) is a fixed point (we may call it the ``off'' node); Eq.\ (\ref{eq:8}) reduces to its first term and \(\Phi(0)=0\).
\item By suitably choosing the half planes---taking advantage of the relative sign in the numerator of the second term in Eq.\ (\ref{eq:8})---another fixed point \(\mathrm{N}:=(\nu_1,\nu_2)^T\) (the ``on'' node) can be induced \footnote{(through an inverse saddle-node bifurcation at the ``on'' location: in one variable, \(\frac{x^2}2-a\,(x-a)\,\theta(x-a)\) resets the slope to zero at \(x=a\))} if \((J\mathrm{N})_k>-\mu_k\), \(k=1,2\) (namely \(j_{11}\nu_1-j_{12}\nu_2>-\mu_1\), \(j_{21}\nu_1-j_{22}\nu_2>-\mu_2\)) with
\[\Phi(\mathrm{N})=\frac{j_{11}j_{21}\nu_1^2-2j_{12}j_{21}\nu_1\nu_2+j_{12}j_{22}\nu_2^2}{2\rho\tau_1\tau_2\det J}+\frac{j_{21}\nu_1\mu_1-j_{12}\nu_2\mu_2}{\rho\tau_1\tau_2\det J}.\]
If \(\mu_1\) is varied (as in \cite{cowan72,boki1992}), equistability is achieved for
\begin{equation}\label{eq:11}
\mu_1=\frac1{2}\left[\frac{j_{12}\nu_2}{j_{21}\nu_1}(j_{21}\nu_1-j_{22}\nu_2+2\,\mu_2)-(j_{11}\nu_1-j_{12}\nu_2)\right].
\end{equation}
The intersection of the cores of the \(s_k(i_k)\) \footnote{(located at the solution \(\frac{j_{22}\mu_1-j_{12}\mu_2}{\det J},\frac{j_{21}\mu_1-j_{11}\mu_2}{\det J}\) of \(J\mathbf{x}+\mathrm{M}=0\))} is a (singular in this limit) saddle point. Figure \ref{fig:1} (b) illustrates this situation for the parameters quoted in the caption (the choice obeys to the fact that global stability makes condition \(j_{21}\nu_1-j_{22}\nu_2>-\mu_2\) rather stringent).
\item As \(\mu_k\to0\), \(k=1,2\), this saddle point moves toward the ``off'' node. After a (direct) saddle-node bifurcation, only the ``on'' node at \(x_k=\nu_k\) remains, since conditions \((J\mathrm{N})_k>-\mu_k\), \(k=1,2\) are better satisfied, see Fig.\ \ref{fig:1} (a).
\end{itemize}
If there is room for some spreading of the core, as seen in Figs.\ \ref{fig:1} (b)--(d), the former result remains valid for whatever analytical form of the response functions. In such a case, the saddle point will be analytical.

In the singular limit \(s_k(i_k):=\nu_k\,\theta(i_k)\) we deal with in this subsection, we can prove rigorously the nonexistence of limit cycles (at least for large \(\mu_k<0\), \(k=1,2\)). The Bendixson--Dulac theorem states that if there exists a \(\mathcal{C}^1\) function \(\Phi(\mathbf{x})\) (called the Dulac function) such that \(\mathrm{div}(\Phi\mathbf{f})\) has the same sign \emph{almost everywhere} \footnote{Everywhere except possibly in a set of measure 0.} in a simply connected region of the plane, then the plane autonomous system \(\dot{\mathbf{x}}=\mathbf{f}(\mathbf{x})\) has no nonconstant periodic solutions lying entirely within the region. Because of Eq.\ (\ref{eq:2}),
\[\mathrm{div}(\Phi\mathbf{f})=\mathbf{f}^\mathrm{T}(\mathbf{x})\nabla\Phi+\Phi\mathrm{div}\,\mathbf{f}=-(\nabla\Phi)^\mathrm{T}Q\,\nabla\Phi+\Phi\mathrm{div}\,\mathbf{f}.\]
Clearly, \(\mathrm{div}\,\mathbf{f}<0\) almost everywhere [i.e.\ except at the cores of the \(s_k(i_k)\)]. For \(\mu_k<0\) (\(k=1,2\)) and large, \(\Phi(\mathbf{x})\) will be essentially the quadratic form in the first term of Eq.\ (\ref{eq:8}), so it meets the conditions to be a Dulac function in a simply connected region of the plane. 
\subsection{Numerically, for a (more) realistic example}\label{ssec:4.2}
\begin{figure}[htbp]
\begin{center}
\includegraphics[width=.45\columnwidth]{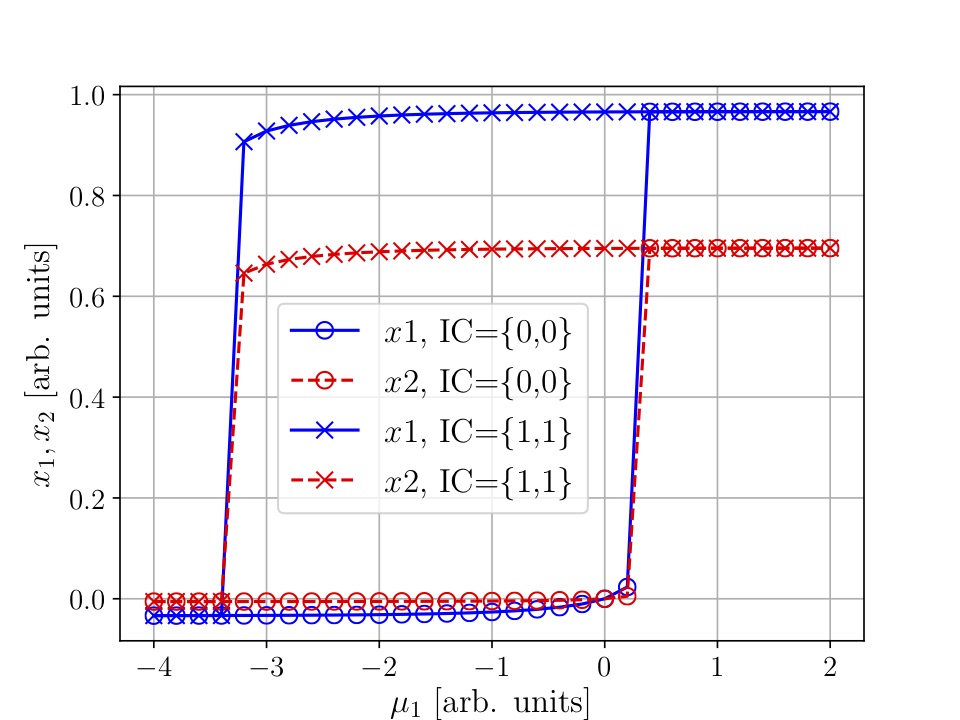}(a)
\includegraphics[width=.45\columnwidth]{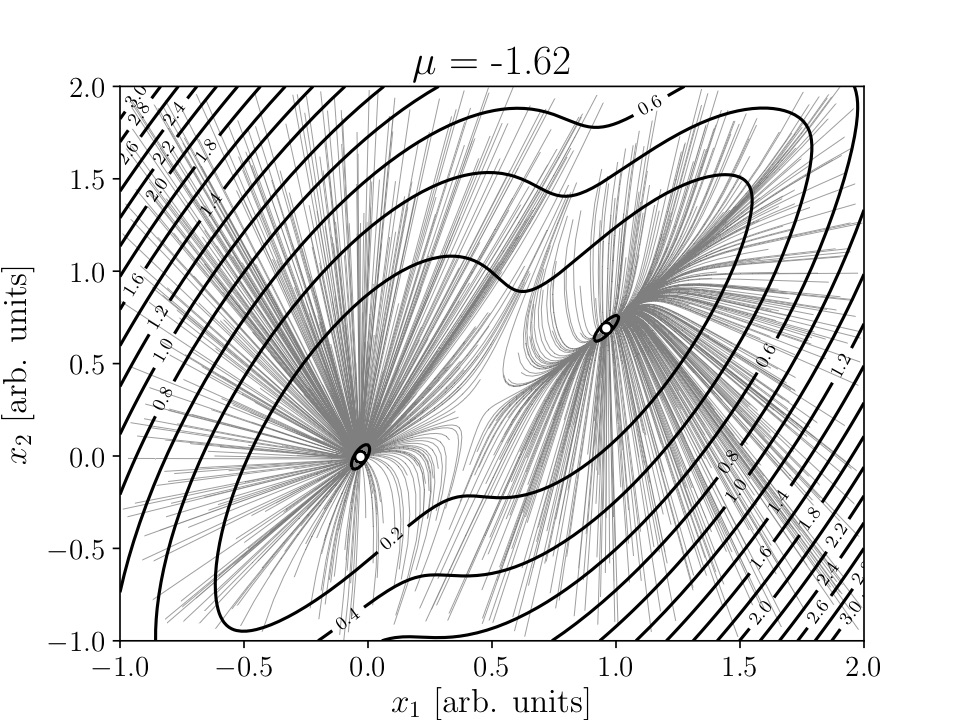}(b)
\includegraphics[width=.45\columnwidth]{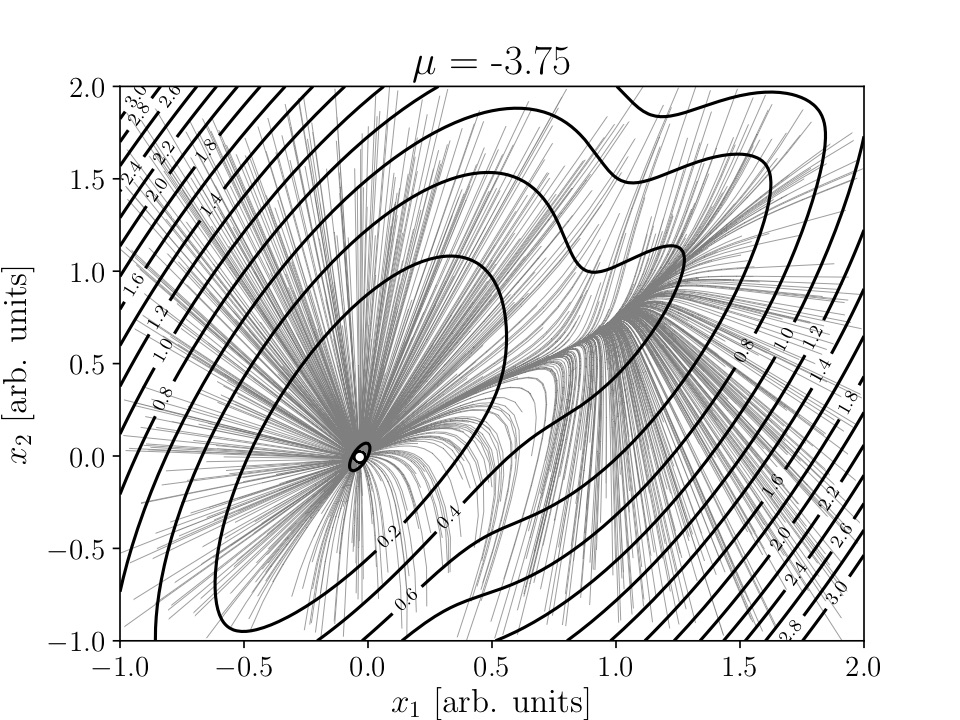}(c)
\includegraphics[width=.45\columnwidth]{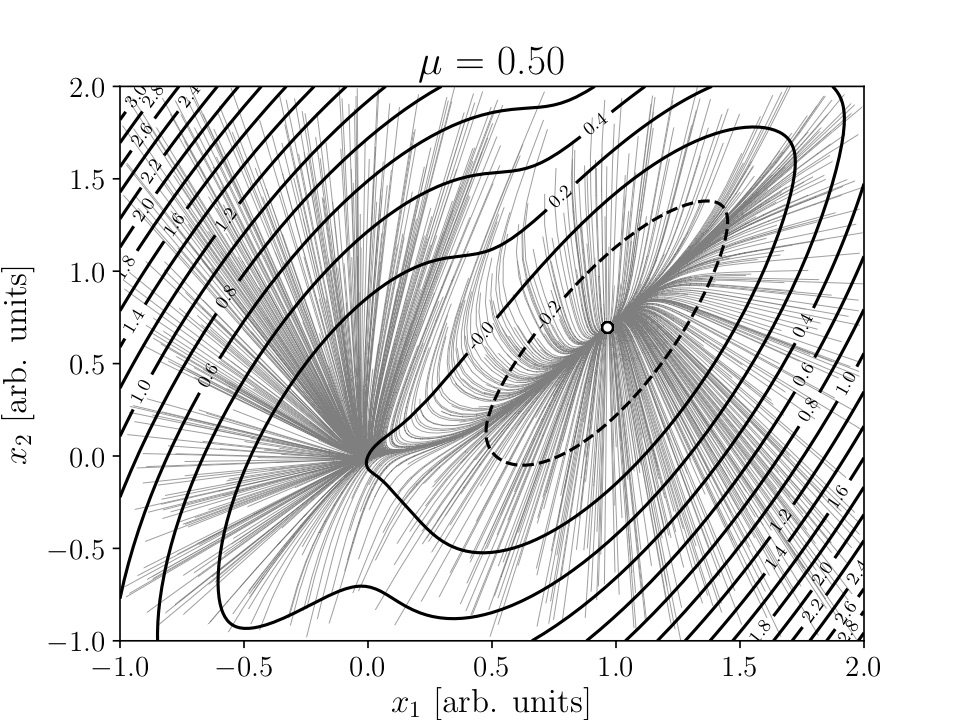}(d)
\caption{(a) Abscissas (solid line) and ordinates (dashed line) of the ``off'' (circle) and ``on'' (cross) nodes of Eqs.\ (\ref{eq:4})--(\ref{eq:5}) (with \(r_1=r_2=0\)) as functions of \(\mu_1\). (b) Trajectories from random initial conditions and contour plot of the NEP---Eq.\ (\ref{eq:8}), with \(S_k( i_k)-S_k(\mu_k)\) given by Eq.\ (\ref{eq:12})---near the equistable case \(\mu_1=-1.7\) (b), and near the ``off'' (c) and ``on'' (d) saddle-node bifurcations. Remaining parameters: \(\rho=1\), \(\tau_1=\tau_2=1\), \(\nu_1=\nu_2=1\), \(\mu_2=0\), \(j_{11}=12\), \(j_{12}=4\), \(j_{21}=13\), \(j_{22}=11\), \(\beta_1=1.2\), \(i_1^0=2.8\), \(\beta_2=1\), and \(i_2^0=4\).}
\label{fig:2}
\end{center}
\end{figure}

For the integrable version (\(r_1=r_2=0\)) of Eqs.\ (\ref{eq:4})--(\ref{eq:5}), it is
\begin{equation}\label{eq:12}
S_k( i_k)-S_k(\mu_k)=\nu_k\left[(i_k-\mu_k)\frac{\exp[\beta_ki_k^0]}{1+\exp[\beta_ki_k^0]}+\beta_k^{-1}\ln\frac{1+\exp[-\beta_k(i_k-i_k^0)]}{1+\exp[-\beta_k(\mu_k-i_k^0)]}\right].
\end{equation}
Here, because of the condition \(s_k(0)=0\), the ``off'' node will move as \(\mu_1\) is varied. Figure \ref{fig:2} considers the integrable version of Figs.\ 4 and 5 (the only ones for which \(\det J<0\)) in \cite{cowan72}. The parameters specified by the authors are \(j_{11}=12\), \(j_{12}=4\), \(j_{21}=13\), \(j_{22}=11\), \(\beta_1=1.2\), \(i_1^0=2.8\), \(\beta_2=1\), \(i_2^0=4\), \(\mu_2=0\). The values of \(\tau_1\) and \(\tau_2\) (as well as \(\nu_1\) and \(\nu_2\), not specified by the authors) have been chosen as 1 throughout \footnote{This has the additional advantage that the flow is purely dissipative, facilitating dynamical conclusions from the landscape.}.

Frame (a), as well as the trajectories from random initial conditions (uniform distribution) in frames (b)--(d), of Figs.\ \ref{fig:1}--\ref{fig:2} are the result of a 4th order Runge--Kutta integration of Eqs.\ (\ref{eq:6}), after 100,000,000 iterations with \(\Delta t=10^{-4}\). In the contour plots of \(\Phi(\mathbf{x})\) of frames (b)--(d) of Fig.\ \ref{fig:2}, \(S_k( i_k)-S_k(\mu_k)\) is given by Eq.\ (\ref{eq:12}). Even though the details differ between Figs.\ \ref{fig:1} and \ref{fig:2}, the structural picture (in particular, the inverse-direct saddle-node mechanism) remains the same.

\section{DIFFERENT RELAXATION TIMES}\label{sec:5}

\begin{figure}[htbp]
\begin{center}
\includegraphics[width=.45\columnwidth]{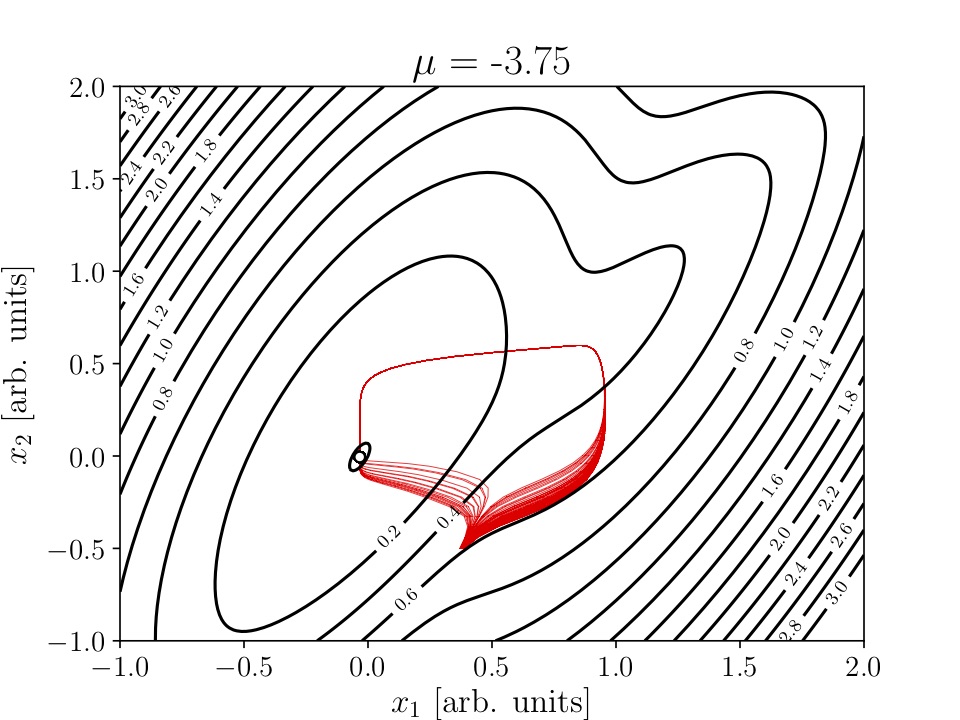}(a)
\includegraphics[width=.45\columnwidth]{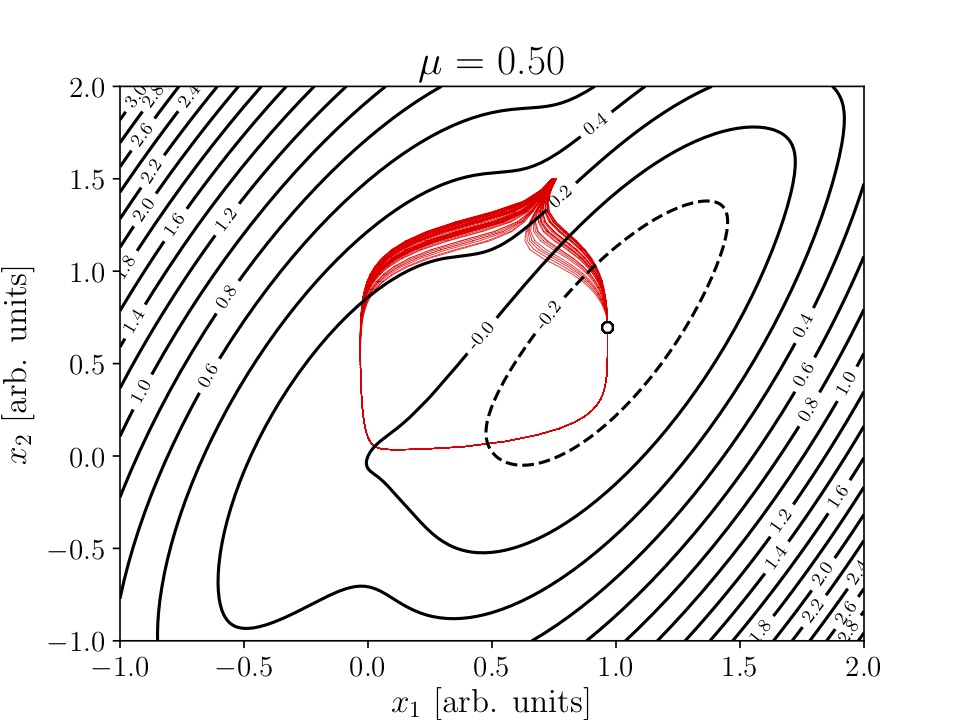}(b)
\caption{Contour plot of the NEP in Figures \ref{fig:2} (c)--(d), together with trajectories (in red) from random initial conditions within suitably selected tiny patches. Parameters: \(\rho=1\), \(\tau_1=0.5\), \(\tau_2=2\), \(\nu_1=\nu_2=1\), \(\mu_2=0\), \(j_{11}=12\), \(j_{12}=4\), \(j_{21}=13\), \(j_{22}=11\), \(\beta_1=1.2\), \(i_1^0=2.8\), \(\beta_2=1\), and \(i_2^0=4\).}
\label{fig:3}
\end{center}
\end{figure}
When \(\tau_1\neq\tau_2\), then
\[\mathbf{r}(\mathbf{x})=\frac{j_{12}j_{21}}{2\rho\det J}\left(\frac1{\tau_1}-\frac1{\tau_2}\right)\left(\begin{array}{c}[-x_1+s_1(i_1)]-\frac{j_{22}}{j_{21}}[-x_2+s_2(i_2)]\\\frac{j_{11}}{j_{12}}[-x_1+s_1(i_1)]-[-x_2+s_2(i_2)]\end{array}\right)\]
and \(\mathbf{d}(\mathbf{x})=\mathbf{f}(\mathbf{x})-\mathbf{r}(\mathbf{x})\). However \(\Phi(\mathbf{x})\) can remain the same, as far as \(\tau_1\tau_2\) does not change. So whereas the contour plots of the NEP in Fig.\ \ref{fig:3} reproduce those of Fig.\ \ref{fig:2} (c)--(d), the displayed set of trajectories (from random initial conditions within suitably selected tiny patches) have \(\mathbf{r}(\mathbf{x})\neq0\) and consequently, many of them perform a large excursion toward the attractor.

Excitable events such as those described here by the NEP, in which the activity of excitatory neurons in the population shows a sharp peak, are know in the computational neuroscience literature as ``population bursts''. In neural network models, they reflect a sudden rise in spiking activity at the level of the whole population (or a significant part of it), in such a way that a high proportion of neurons in the network fire at least one action potential during a short time window. The Wilson-Cowan model captures this phenomenon as a transient peak of activity that is later shut down by inhibition. But for more realistic models, additional biophysical mechanisms (such as the refractory period of neurons or short-term adaptation) are also involved \cite{tsum2000}. Population bursts have several computational uses; for example, they can be used to transmit temporally precise information to other brain areas, even in the presence of noise or heterogeneity \cite{Mejias2012}.

\section{CONCLUSIONS}\label{sec:7}
\emph{Rate}- (also called \emph{neural mass}-) models have been a useful approach to neural networks for half a century. Today, their simplicity (not short of comprehensivity) makes them ideal to fulfill the node dynamics in, for instance, connectome-based brain networks. So the availability of an ``energy function'' for rate models is expected to be welcome news.

Dynamical systems of the form given by Eq.\ (\ref{eq:6}) admit a NEP \emph{regardless} of the functional forms of the nonlinear single-variable functions \(s_k(i_k)\). Throughout this work, the latter are assumed to have the same functional form, of sigmoidal shape. But neither condition is necessary to satisfy the integrability condition.

A crucial observation about rate models---even the one put forward by Wilson and Cowan \cite{cowan72}, and given by Eqs.\ (\ref{eq:4})--(\ref{eq:5})---is that they are \emph{asymptotically linear}, so their eventual NEP can be \emph{at most quadratic}. Then in principle, \emph{global stability} rules out some coupling configurations. Obviously, this requirement can be relaxed if the rate model fulfills the node dynamics of a neural network, for what matters in that case is the network's global stability.

The here obtained NEP provides a more quantitative intuition on the phenomenon of bistability, that has been naturally found in real neural systems. Neural bistability underlies e.g. the persistent activity which is commonly found in neurons of the prefrontal cortex, a mechanism that is thought to maintain information during working memory tasks \cite{Funahashi1989,wang2001}. In the presence of neural noise and other adaptation mechanisms, bistability is also a useful hypothesis to explain slow irregular dynamics or `up' and `down' dynamics, also observed across cortex and modeled using bistable dynamics \cite{McCormick2005,Tsodyks2006,Mejias2010}. 

Finally, it is worth mentioning that the here obtained NEP---valid as argued for generic transfer functions \(s_k(i_k)\)---opens the door to the potential use of more generic rate models in the field of \emph{artificial} neural networks and \emph{deep learning}. By identifying the NEP with the cost function to be minimized, gradient descent algorithms can be used to train networks of generic Wilson--Cowan units for different tasks. This implies that more realistic and less computationally expensive neural population models can be trained and used for behavioral tasks, a topic that has gathered attention recently \cite{Mante2013,Song2016}.

\section*{Author Contributions}

All the listed authors have made substantial and direct intellectual contribution to this work, and approve its publication. In particular, JFM proposed the idea, contributed preliminary analytical calculations, and provided key information from the field of neuroscience.

\section*{Funding}
RRD acknowledges support from UNMdP \texttt{http://dx.doi.org/10.13039/501100007070}, under Grant 15/E779--EXA826/17. NM acknowledges a Doctoral Fellowship from CONICET \texttt{http://dx.doi.org/10.13039/501100002923}.

\section*{Acknowledgments}
JFM thanks warmly A. Longtin for his support during the early stages of this work. JID and HSW thank respectively IFIMAR and IFISC for their hospitality. NM thanks HSW and IFISC for their hospitality during his research visit.

\begin{thebibliography}{}

\bibitem[Amit and Brunel, 1997]{Amit1997}
Amit, D.~J. and Brunel, N. (1997).
\newblock Model of global spontaneous activity and local structured activity
  during delay periods in the cerebral cortex.
\newblock {\em Cerebral Cortex}, 7(3):237--252.

\bibitem[Ao, 2004]{ao04}
Ao, P. (2004).
\newblock Potential in stochastic differential equations: novel construction.
\newblock {\em J. Phys. A}, 37:L25--L30.

\bibitem[Borisyuk and Kirillov, 1992]{boki1992}
Borisyuk, R.~M. and Kirillov, A.~B. (1992).
\newblock Bifurcation analysis of a neural network model.
\newblock {\em Biol. Cybern.}, 66:319--325.

\bibitem[Brunel, 2000]{brunel00}
Brunel, N. (2000).
\newblock Dynamics of sparsely connected networks of excitatory and inhibitory
  spiking neurons.
\newblock {\em J. Comp. Neurosci.}, 8:183--208.

\bibitem[Descalzi and Graham, 1992]{degr}
Descalzi, O. and Graham, R. (1992).
\newblock Gradient expansion of the nonequilibrium potential for the
  supercritical {Ginzburg--Landau} equation.
\newblock {\em Phys. Lett. A}, 170:84.

\bibitem[Funahashi et~al., 1989]{Funahashi1989}
Funahashi, S., Bruce, C.~J., and Goldman-Rakic, P.~S. (1989).
\newblock Mnemonic coding of visual space in the monkey9s dorsolateral
  prefrontal cortex.
\newblock {\em J. Neurophysiol.}, 61:331--349.

\bibitem[Gardiner, 2004]{gard04}
Gardiner, C.~W. (2004).
\newblock {\em Handbook of Stochastic Methods for Physics, Chemistry and the
  Natural Sciences}.
\newblock Springer, Berlin.

\bibitem[Graham and T\'el, 1990]{grte90}
Graham, R. and T\'el, T. (1990).
\newblock Steady-state ensemble for the complex {Ginzburg--Landau} equation
  with weak noise.
\newblock {\em Phys. Rev. A}, 42:4661.

\bibitem[{Graham R ``Weak noise limit and nonequilibrium potentials of
  dissipative dynamical systems'' In Tirapegui E, Villarroel D}, 1987]{graham}
{Graham R ``Weak noise limit and nonequilibrium potentials of dissipative
  dynamical systems'' In Tirapegui E, Villarroel D}, editor (1987).
\newblock {\em Instabilities and Nonequilibrium Structures}.
\newblock Reidel, Dordrecht.

\bibitem[Holcman and Tsodyks, 2006]{Tsodyks2006}
Holcman, D. and Tsodyks, M. (2006).
\newblock The emergence of up and down states in cortical networks.
\newblock {\em PLoS Comput. Biol.}, 2:174--181.

\bibitem[Iz\'us et~al., 1998]{wio98}
Iz\'us, G.~G., Deza, R.~R., and Wio, H.~S. (1998).
\newblock Exact nonequilibrium potential for the {FitzHugh--Nagumo} model in
  the excitable and bistable regimes.
\newblock {\em Phys. Rev. E}, 58:93--98.

\bibitem[Iz\'us et~al., 1999]{wio99}
Iz\'us, G.~G., Deza, R.~R., and Wio, H.~S. (1999).
\newblock Critical slowing-down in the {FitzHugh--Nagumo} model: A
  non-equilibrium potential approach.
\newblock {\em Comp. Phys. Comm.}, 121-122:406--407.

\bibitem[Iz\'us et~al., 2009]{Deza2009}
Iz\'us, G.~G., S\'anchez, A.~D., and Deza, R.~R. (2009).
\newblock Noise-driven synchronization of a {FitzHugh--Nagumo} ring with
  phase-repulsive coupling: A perspective from the system's nonequilibrium
  potential.
\newblock {\em Physica A}, 388(6):967--976.

\bibitem[Kim and Wang, 2007]{kiwa07}
Kim, K.-Y. and Wang, J. (2007).
\newblock Potential energy landscape and robustness of a gene regulatory
  network: Toggle switch.
\newblock {\em PLoS Comp. Biol.}, 3:e60.

\bibitem[Kirkpatrick et~al., 1983]{kiea83}
Kirkpatrick, S., Gelatt, C.~D., and Vecchi, M.~P. (1983).
\newblock Optimization by simulated annealing.
\newblock {\em Science}, 220:671--680.

\bibitem[Langevin, 1908]{lang}
Langevin, P. (1908).
\newblock Sur la th\'eorie du mouvement {B}rownien.
\newblock {\em C. R. Acad. Sci. (Paris)}, 146:530.

\bibitem[Li et~al., 2011]{liea11}
Li, C., Wang, E., and Wang, J. (2011).
\newblock Potential landscape and probabilistic flux of a predator prey
  network.
\newblock {\em PLoS Comp. Biol.}, 6:e17888.

\bibitem[Lyapunov, 1892]{lyap}
Lyapunov, A.~M. (1892).
\newblock {\em The general problem of the stability of motion \emph{(in
  russian)}}.
\newblock Mathematical Society of Kharkov, Kharkiv, Ukraine.
\newblock english transl.: Int. J. Control \textbf{55} (1992), 531--772.

\bibitem[Mante et~al., 2013]{Mante2013}
Mante, V., Sussillo, D., Shenoy, K.~V., and Newsome, W.~T. (2013).
\newblock Context-dependent computation by recurrent dynamics in prefrontal
  cortex.
\newblock {\em Nature (London)}, 503:78--84.

\bibitem[McCormick, 2005]{McCormick2005}
McCormick, D.~A. (2005).
\newblock Neuronal networks: flip-flops in the brain.
\newblock {\em Curr. Biol.}, 15:R294--R296.

\bibitem[Mej\'{\i}as et~al., 2010]{Mejias2010}
Mej\'{\i}as, J.~F., Kappen, H.~J., and Torres, J.~J. (2010).
\newblock Irregular dynamics in up and down cortical states.
\newblock {\em PLoS ONE}, 5:e13651.

\bibitem[Mej\'{\i}as and Longtin, 2012]{Mejias2012}
Mej\'{\i}as, J.~F. and Longtin, A. (2012).
\newblock Optimal heterogeneity for coding in spiking neural networks.
\newblock {\em Phys. Rev. Lett.}, 108:228102.

\bibitem[Mej\'{\i}as and Longtin, 2014]{Mejias2014b}
Mej\'{\i}as, J.~F. and Longtin, A. (2014).
\newblock Differential effects of excitatory and inhibitory heterogeneity on
  the gain and asynchronous state of sparse cortical networks.
\newblock {\em Frontiers Comp. Neurosci.}, 8:107.

\bibitem[Montagne et~al., 1996]{moea}
Montagne, R., Hern\'andez-Garc\'{\i}a, E., and {San Miguel}, M. (1996).
\newblock Numerical study of a {Lyapunov} functional for the complex
  {Ginzburg--Landau} equation.
\newblock {\em Physica D}, 96:47.

\bibitem[Risken, 1996]{risken}
Risken, H. (1996).
\newblock {\em The Fokker-Planck Equation: Methods of solution and
  applications\emph{, 2nd ed.}}
\newblock Springer, Berlin.

\bibitem[S\'anchez et~al., 2014]{Deza2014}
S\'anchez, A., Iz\'us, G., dell'Erba, M., and Deza, R. (2014).
\newblock A reduced gradient description of stochastic-resonant spatiotemporal
  patterns in a {FitzHugh--Nagumo} ring with electric inhibitory coupling.
\newblock {\em Phys. Lett. A}, 378:1579--1583.

\bibitem[S\'anchez and Iz\'us, 2010]{Izus2010}
S\'anchez, A.~D. and Iz\'us, G.~G. (2010).
\newblock Nonequilibrium potential for arbitrary-connected networks of
  {FitzHugh--Nagumo} elements.
\newblock {\em Physica A}, 389(9):1931--1944.

\bibitem[Song et~al., 2016]{Song2016}
Song, H.~F., Yang, G.~R., and Wang, X.~J. (2016).
\newblock Training excitatory-inhibitory recurrent neural networks for
  cognitive tasks: A simple and flexible framework.
\newblock {\em PLoS Comp. Biol.}, 12:e1004792.

\bibitem[Tsodyks et~al., 2000]{tsum2000}
Tsodyks, M., Uziel, A., and Markram, H. (2000).
\newblock Synchrony generation in recurrent networks with frequency-dependent
  synapses.
\newblock {\em J. Neurosci.}, 20:RC50.

\bibitem[{van Kampen}, 1990]{vaka90}
{van Kampen}, N.~G. (1990).
\newblock {\em Stochastic processes in physics and chemistry}.
\newblock North-Holland, Amsterdam.

\bibitem[Wang et~al., 2006]{waea06}
Wang, J., Huang, B., Xia, X., and Sun, Z. (2006).
\newblock Funneled landscape leads to robustness of cell networks: Yeast cell
  cycle.
\newblock {\em PLoS Comp. Biol.}, 2:e147.

\bibitem[Wang et~al., 2012]{waea12}
Wang, J., Oliveira, R.~J., Chu, X., Whitford, P.~C., Chahine, J., Han, W.,
  Wang, E., Onuchic, J.~N., and Leite, V.~B. (2012).
\newblock Topography of funneled landscapes determines the thermodynamics and
  kinetics of protein folding.
\newblock {\em Proc. Nat. Acad. Sci. USA}, 109:15763--15768.

\bibitem[Wang et~al., 2010]{waea10}
Wang, J., Xu, L., Wang, E., and Huang, S. (2010).
\newblock The potential landscape of genetic circuits imposes the arrow of time
  in stem cell differentiation.
\newblock {\em Biophys. J.}, 99:29--39.

\bibitem[Wang, 2001]{wang2001}
Wang, X.-J. (2001).
\newblock Synaptic reverberation underlying mnemonic persistent activity.
\newblock {\em Trends Neurosci.}, 24:455--463.

\bibitem[Wang, 2002]{wang2002}
Wang, X.-J. (2002).
\newblock Probabilistic decision making by slow reverberation in cortical
  circuits.
\newblock {\em Neuron}, 36:955--968.

\bibitem[Wilson and Cowan, 1972]{cowan72}
Wilson, H.~R. and Cowan, J.~D. (1972).
\newblock Excitatory and inhibitory interactions in localized populations of
  model neurons.
\newblock {\em Biophys. J.}, 12:1--24.

\bibitem[Wio et~al., 2002]{wvhb2002}
Wio, H., {Von Haeften}, B., and Bouzat, S. (2002).
\newblock Stochastic resonance in spatially extended systems: the role of far
  from equilibrium potentials.
\newblock {\em Physica A}, 306:140--156.

\bibitem[Wio and Deza, 2007]{epjst146:111}
Wio, H.~S. and Deza, R.~R. (2007).
\newblock Aspects of stochastic resonance in reaction--diffusion systems: The
  nonequilibrium-potential approach.
\newblock {\em Eur. Phys. J. Special Topics}, 146:111--126.

\bibitem[Wio et~al., 2012]{wdl}
Wio, H.~S., Deza, R.~R., and L\'opez, J.~M. (2012).
\newblock {\em An Introduction to Stochastic Processes and Nonequilibrium
  Statistical Physics, \emph{revised edition}}.
\newblock World Scientific, Singapore.

\bibitem[{Wio HS ``Nonequilibrium potential in reaction--diffusion systems'' In
  Garrido P, Marro J}, 1997]{Wio}
{Wio HS ``Nonequilibrium potential in reaction--diffusion systems'' In Garrido
  P, Marro J}, editor (1997).
\newblock {\em 4th Granada Seminar in Computational Physics}.
\newblock Springer, Dordrecht.

\bibitem[Wong and Wang, 2006]{wong2006}
Wong, K. and Wang, X.-J. (2006).
\newblock A recurrent network mechanism of time integration in perceptual
  decisions.
\newblock {\em J. Neurosci.}, 26:1314--1328.

\bibitem[Wu and Wang, 2013]{wuwa13}
Wu, W. and Wang, J. (2013).
\newblock Landscape framework and global stability for stochastic reaction
  diffusion and general spatially extended systems with intrinsic fluctuations.
\newblock {\em J. Phys. Chem. B}, 117:12908--12934.

\bibitem[Yan et~al., 2013]{Yan2013}
Yan, H., Zhao, L., Hu, L., Wang, X., Wang, E., and Wang, J. (2013).
\newblock Nonequilibrium landscape theory of neural networks.
\newblock {\em Proc. Nat. Acad. Sci. USA}, 110 (45):E4185--94.

\bibitem[Zhang et~al., 2012]{zhea12}
Zhang, F., Xu, L., Zhang, K., Wang, E., and Wang, J. (2012).
\newblock The potential and flux landscape theory of evolution.
\newblock {\em J. Chem. Phys.}, 137:065102.

\end{thebibliography}

\end{document}